\documentclass[11pt,fleqn]{article}
\usepackage{amssymb}

\usepackage{amsmath}

\usepackage{amsmath,amssymb,amsthm}

\begin{document}

\begin{abstract}
New concept of conditional differential invariant is discussed that would
allow description of equations invariant with respect to an operator under a
certain condition. Example of conditional invariants of the projective
operator is presented.
\end{abstract}

\begin{center}
{\Large \textbf{Differential Invariants and Construction\\[0pt]
of Conditionally Invariant Equations }}

\vskip 3pt {\large \textbf{Irina YEHORCHENKO }} \vskip 6pt Institute of
Mathematics of NAS of Ukraine, 3 Tereshchenkivska Str., Kyiv 4, Ukraine\\[0pt%
]
E-mail: iyegorch@imath.kiev.ua
\end{center}

\section{Introduction}

Importance of investigation of symmetry properties of differential equations
is well-established in mathematical physics. Classical methods for studying
symmetry properties and their utilisation for finding solutions of partial
differential equations were originated in the papers by S.~Lie, and
developed by modern authors (see e.g.\ \cite
{yehorchenko:Ovs-eng,yehorchenko:Olver1,yehorchenko:Bluman Kumei
book,yehorchenko:FSS}).

We start our consideration from some symmetry properties and solutions of
the nonlinear wave equation
\begin{equation}  \label{yehorchenko:nl wave eq}
\Box u = F (u, u^*)
\end{equation}
for the complex-valued function $u = u(x_0,x_1,\ldots,x_n)$, $x_0 =t$ is the
time variable, $x_1,\ldots,x_n$ are $n$ space variables. $F$ is some
function. $\Box u $ is the d'Alembert operator
\begin{equation}
\Box u = - \frac{\partial^2 u}{\partial x^2_0} + \frac{\partial^2 u }{%
\partial x^2_1} +\cdots + {\frac{\partial^2 u }{\partial x^2_n}}.
\end{equation}

It is well-known that the equation~(\ref{yehorchenko:nl wave eq}) may be
reduced to a nonlinear Schr\"odinger equation with the number of space
dimensions smaller by 1, when the nonlinearity $F$ has a special form $F=u
f(|u|)$, where $|u|=(uu^*)^{1/2}$, an asterisk designates complex
conjugation.

Further we are trying to generalise this relation between the nonlinear wave
equation and the nonlinear Schr\"odinger equation into a relation between
differential invariants of the respective invariance algebrae, and introduce
new concepts of the reduction of fundamental sets of differential invariants
and of conditional differential invariants. Conditional differential
invariants may be utilised to describe conditionally invariant equations
under certain operators and with the certain conditions, in the same manner
as absolute differential invariants of a Lie algebra may be utilised for
description of all equations invariant under this algebra.

The concept of non-classical, or conditional symmetry, originated in its
various facets in the papers \cite{yehorchenko:Bluman Cole,
yehorchenko:Olver Rosenau,yehorchenko:F Tsyfra,yehorchenko:F Zhdanov
cond,yehorchenko:Clarkson Kruskal,yehorchenko:Levi Winternitz} and later by
numerous authors was developed into the theory and a number of algorithms
for studying symmetry properties of equations of mathematical physics and
for construction of their exact solutions. Here we will use the following
definition of the conditional symmetry:

\textbf{Definition 1} The equation $F(x,u,{\mathop {u}\limits_1},\ldots ,{%
\mathop
{u}\limits_l})=0$ where ${\mathop {u}\limits_k}$ is the set of all $k\,%
\mathrm{th}$-order partial derivatives of the function $u=(u^1,u^2,\ldots
,u^m)$, is called conditionally invariant under the ope\-rator
\begin{equation}
Q=\xi ^i(x,u)\partial _{x_i}+\eta ^r(x,u)\partial _{u^r}
\end{equation}
if there is an additional condition
\begin{equation}
G(x,u,{\mathop {u}\limits_1},\ldots ,{\mathop {u}\limits_{l_1}})=0,
\label{yehorchenko:G=0}
\end{equation}
\noindent such that the system of two equations $F=0$, $G=O$ is invariant
under the operator $Q$.

If (\ref{yehorchenko:G=0}) has the form $G=Qu$, then the equation $F=0$ is
called $Q$-conditionally invariant under the operator $Q$.

\section{Differential invariants and description of invariant equations}

Differential invariants of Lie algebrae present a powerful tool for studying
partial differential equations and construction of their solutions \cite
{yehorchenko:Lie,yehorchenko:Tresse,yehorchenko:Olver2}.

Now we will present some basic definitions that we will further generalise.
For the purpose of these definitions we deal with Lie algebrae consisting of
the infinitesimal operators
\begin{equation}  \label{yehorchenko:X}
X = \xi^i (x, u) \partial_{x_i} + \eta^r (x, u) \partial_{u^r}.
\end{equation}
Here $x=(x_1,x_2,\ldots,x_n)$, $u=(u^1,u^2,\ldots,u^m)$.

\textbf{Definition 2} The function $F=F(x,u,{\mathop {u}\limits_1},\ldots ,{%
\mathop {u}\limits_l}),$ is called a differential invariant for the Lie
algebra $L$ with basis elements $X_i$ of the form~(\ref{yehorchenko:X}) $%
(L=\langle X_i\rangle )$ if it is an invariant of the $l\,\mathrm{th}$
prolongation of this algebra:
\begin{equation}
{\mathop {X}\limits_l}_sF(x,u,{\mathop {u}\limits_1},...,{\mathop {u}%
\limits_l})=\lambda _s(x,u,{\mathop {u}\limits_1}...,{\mathop {u}\limits_l}%
)F,
\end{equation}

\noindent where the $\lambda _s$ are some functions; when $\lambda _i=0,F$
is called an absolute invariant; when $\lambda _i\ne 0$, it is a relative
invariant.

Further when writing ``differential invariant'' we would imply ``absolute
differential invariant''.

\textbf{Definition 3} A maximal set of functionally independent invariants
of order $r\leq l$ of the Lie algebra $L$ is called a functional basis of
the $l\,\mathrm{th}$-order differential invariants for the algebra $L$.

While writing out lists of invariants we shall use the following
designations
\begin{gather}
u_a \equiv \frac{\partial u}{\partial x_a}, \qquad u_{ab} \equiv \frac{%
\partial^2 u}{\partial x_a \partial x_b}, \qquad S_k (u_{ab}) \equiv
u_{a_1a_2} u_{a_2a_3} \cdots u_{a_{k-1} a_k} u_{a_k a_1},  \nonumber \\
S_{jk} (u_{ab}, v_{ab}) \equiv u_{a_1a_2} \cdots u_{a_{j-1} a_j} v_{a_j
a_{j+1}} \cdots v_{a_k a_1},  \nonumber \\
R_{k}(u_{a},u_{ab})\equiv u_{a_1}u_{a_k}u_{a_1 a_2}u_{a_2 a_3 }\cdots
u_{a_{k-1} a_k}.  \label{yehorchenko:SR}
\end{gather}

In all the lists of invariants $j$ takes the values from 0 to $k$. We shall
not discern the upper and lower indices with respect to summation: for all
Latin indices $x_a x_a \equiv x_a x^a \equiv x^a x_a =x_1^2 +x_2^2 +\cdots +
x_n^2$.

Fundamental bases of differential invariants for the standard scalar
representations of the Poincar\'e and Galilei algebra of the types~(\ref
{yehorchenko:poin}), (\ref{yehorchenko:Galilei1}) were found in~\cite
{yehorchenko:F Ye DifInvs}. Fundamental bases of differential invariants
allow describing all equations invariant under the respective Lie algebrae.

Construction of conditional differential invariants would allow describing
all equations, conditionally invariant with respect to certain operators
under certain conditions.

\textbf{Definition 4} $F=F(x,u,{\mathop {u}\limits_1},\ldots ,{\mathop
{u}\limits_l})$ is called a conditional differential invariant for the
operator with $X$ of the form~(\ref{yehorchenko:X}) if under the condition
\begin{gather}
G(x,u,{\mathop {u}\limits_1},\ldots ,{\mathop {u}\limits_{l_1}})=0,
\label{yehorchenko:Gcond} \\
{\mathop {X}\limits_{l_{\max }}}F(x,u,{\mathop {u}\limits_1},\ldots ,{%
\mathop {u}\limits_l})=0,\qquad {\mathop {X}\limits_{l_{\max }}}G(x,u,{%
\mathop {u}\limits_1},\ldots ,{\mathop {u}\limits_{l_1}})=0,
\label{yehorchenko:Xcond}
\end{gather}
${\mathop {X}\limits_{l_{\max }}}$ being the $l_{\max }\,\mathrm{th}$
prolongation of the operator $X$. The order of the prolongation $l_{\max }={%
\max }(l,l_1)$.

\section{Nonlinear wave equation, nonlinear Schr\"odinger equation\newline
and relation between their symmetries}

The Galilei algebra for $n-1$ space dimensions is a subalgebra of the
Poincar\'e algebra for $n$ space dimensions (see e.g.~\cite
{yehorchenko:Gomis}) and references therein), and this fact allows reduction
of the nonlinear wave equation~(\ref{yehorchenko:nl wave eq}) to the
Schr\"odinger equation. We will consider the nonlinear wave equations for
three space variables, and its symmetry properties in relation to the
symmetry properties of the nonlinear Schr\"odinger equation for two space
variables. However, all the results can be easily generalised for arbitrary
number of space dimensions.

Reduction of the nonlinear wave equation~(\ref{yehorchenko:nl wave eq}) to
the Schr\"odinger equation can be performed by means of the ansatz
\begin{equation}  \label{yehorchenko:ansatz}
u = \exp((-im/2) (x_0+x_3)) \Phi (x_0-x_3,x_1,x_2).
\end{equation}

Substitution of the expression~(\ref{yehorchenko:ansatz}) into~(\ref
{yehorchenko:nl wave eq}) gives the equation $\exp(({-im/2}) (x_0+x_3))
(2im\Phi_{\tau} + \Phi_{11}+ \Phi_{22})= F(u,u^*).$ Here we adopted the
following notations: $\tau = x_0+x_3$ is the new time variable, $%
\Phi_{\tau}= \frac{\partial \Phi}{\partial \tau}$, $\Phi_{a}= \frac{\partial
\Phi}{\partial x_a }$, $\Phi_{ab}=\frac{\partial^2 \Phi}{\partial x_a
\partial x_b }.$

Further on we adopt the convention that summation is implied over the
repeated indices. If not stated otherwise, small Latin indices run from 1 to
2.

If the nonlinearity in the equation~(\ref{yehorchenko:nl wave eq}) has the
form $F=uf(|u|)$, then it reduces to the Schr\"odinger equation
\begin{equation}  \label{yehorchenko:Schr2}
2im\Phi_{\tau} + \Phi_{11}+ \Phi_{22}= \Phi f (|\Phi|).
\end{equation}

Such reduction allowed construction of numerous new solutions for the
nonlinear wave equation by means of the solutions of a nonlinear
Schr\"odinger equation \cite{yehorchenko:Basarab2,yehorchenko:Basarab1}. We
show that this reduction allowed also to describe additional symmetry
properties for the equation~(\ref{yehorchenko:nl wave eq}), related to the
symmetry properties of the equation~(\ref{yehorchenko:Schr2}).

Lie symmetry of the equation~(\ref{yehorchenko:Schr2}) was described in \cite
{yehorchenko:FMosk,yehorchenko:W1}. With an arbitrary function $f$ it is
invariant under the Galilei algebra with basis operators
\begin{gather}
\partial_\tau =\frac{\partial}{\partial \tau}, \qquad \partial_a = \frac{%
\partial}{\partial x_a}, \qquad J_{12} =x_1\partial_2 - x_2 \partial_1,
\nonumber \\
G_a =t\partial_a +ix_a (\Phi \partial_\Phi - \Phi^* \partial_{\Phi^*}) \quad
(a=1,2), \qquad J =(\Phi \partial_\Phi - \Phi^* \partial_{\Phi^*}).
\label{yehorchenko:Galilei1}
\end{gather}

When $f = \lambda |u|^2$, where $\lambda$ is an arbitrary constant, the
equation~(\ref{yehorchenko:Schr2}) is invariant under the extended Galilei
algebra that contains besides the operators ~(\ref{yehorchenko:Galilei1})
also the dilation operator
\begin{equation}  \label{yehorchenko:D}
D = 2\tau \partial_\tau + x_a \partial_a - I,
\end{equation}
where $I = \Phi\partial_\Phi + \Phi^* \partial_{\Phi^*}$, and the projective
operator
\begin{equation}  \label{yehorchenko:A}
A= \tau^2 \partial_\tau + \tau x_a \partial_a +{\frac{im }{2}}x_a x_a J-
\tau I.
\end{equation}
Lie reductions and families of exact solutions for multidimensional
nonlinear Schr\"odinger equations were found at \cite{yehorchenko:FSerov
Schr,yehorchenko:W1,yehorchenko:W2,yehorchenko:W3,yehorchenko:W4,yehorchenko:W5}%
. Note that the ansatz~(\ref{yehorchenko:ansatz}) is the general solution of
the equation
\begin{equation}  \label{yehorchenko:condition}
u_0 + u_3 + imu=0.
\end{equation}

We can regard the equation~(\ref{yehorchenko:condition}) as the additional
condition imposed on the nonlinear wave equation with the nonlinearity $%
F=\lambda u|u|^2$. Solution of the resulting system
\begin{equation}  \label{yehorchenko:nl wave2}
\Box u = \lambda u|u|^2,
\end{equation}
with the equation~(\ref{yehorchenko:condition}) would allow to reduce number
of independent variables by one, and obtain the same reduced equation,
invariant under the extended Galilei algebra with the projective operator.
This allows establishing conditional invariance of the nonlinear wave
equation~(\ref{yehorchenko:nl wave2}) under the projective operator. It is
well-known that it is not invariant under this operator in the Lie sense.

The maximal invariance algebra of the equation~(\ref{yehorchenko:nl wave eq}%
) that may be found according to the Lie algorithm (see e.g.\ \cite
{yehorchenko:Ovs-eng,yehorchenko:Olver1,yehorchenko:Bluman Kumei
book,yehorchenko:FSS}) is defined by the following basis operators:
\begin{equation}  \label{yehorchenko:poin}
p_{\mu} = ig_{\mu \nu} {\frac{\partial }{\partial x_{\nu}}}, \qquad J_{\mu
\nu} = x_{\mu} p_{\nu} -x_{\nu} p_{\mu},
\end{equation}
where $\mu, \nu$ take the values $0,1,\ldots,3$; the summation is implied
over the repeated indices (if they are small Greek letters) in the following
way: $x_{\nu } x_{\nu} \equiv x_{\nu} x^{\nu} \equiv x^{\nu} x_{\nu} =x_0^2
-x_1^2 - \cdots - x_n^2$, $g_{\mu \nu }= \mathrm{diag}\,(1,-1,\ldots,-1). $

However, summation for all derivatives of the function $u$ is assumed as
follows: $u_{\nu } u_{\nu} \equiv u_{\nu} u^{\nu} \equiv u^{\nu} u_{\nu}
=-u_0^2 +u_1^2 + \cdots + u_n^2$.

Unlike the standard convention on summation of the repeated upper and lower
indices we consider $x_{\nu}$ and $x^{\nu}$ equal with respect to summation
not to mix signs of derivatives and numbers of functions.

\textbf{Theorem 1} The nonlinear wave equation~(\ref{yehorchenko:nl wave2})
is conditionally invariant with the condition~(\ref{yehorchenko:condition})
under the projective operator
\begin{gather}
A_1=\frac 12(x_0-x_3)^2(\partial _0-\partial _3)+(x_0-x_3)(x_1\partial
_1+x_2\partial _2)  \nonumber \\
\qquad {}+\frac{imx^2}2(u\partial _u-u^{*}\partial _{u^{*}})+\frac{n-1}%
2(x_0-x_3)(u\partial _u+u^{*}\partial _{u^{*}}).  \label{yehorchenko:A1}
\end{gather}

To prove Theorem 1 it is sufficient to show that the system~(\ref
{yehorchenko:nl wave2}), (\ref{yehorchenko:condition}) is invariant under
the operator~(\ref{yehorchenko:A1}) by means of the classical Lie algorithm.

Our further study aims at construction of other Poincar\'e-invariant
equations possessing the same conditional invariance property.

\section{Example: construction of conditional differential invariants}

\noindent Now we adduce fundamental bases of differential invariants that
will be utilised for construction of our example of conditional differential
invariants.

First we present a functional basis of differential invariants for the
Poincar\'e algebra~(\ref{yehorchenko:poin}) of the second order for the
complex-valued scalar function $u = u (x_0,x_1,\ldots,x_3).$ It consists of
24 invariants
\begin{equation}  \label{yehorchenko:poin inv}
u^r,\qquad R_k\left(u^r_{\mu}, u^1_{\mu\nu} \right),\qquad S_{jk}
\left(u^r_{\mu\nu}, u^1_{\mu\nu}\right).
\end{equation}

In~(\ref{yehorchenko:poin inv}) everywhere $k=1,\ldots,4$; $j=0,\ldots,k$. A
functional basis of differential invariants for the Galilei algebra~(\ref
{yehorchenko:Galilei1}), mass $m \ne 0$, of the second order for the
complex-valued scalar function $\Phi = \Phi (\tau,x_1,\ldots,x_2)$ consists
of 16 invariants.

For simplification of the expressions for differential invariants we
introduced the following notations:
\[
\Phi = \exp\phi, \qquad \mathrm{Im}\, \Phi = \arctan \frac{\mathrm{Re}\,\phi%
}{\mathrm{Im}\,\phi}\,.
\]

The elements of the functional basis may be chosen as follows:
\begin{gather}
\phi + \phi^*, \quad M_1 = 2im \phi_{t} + \phi_{a} \phi_{a},\quad M_1^* ,
\quad M_2 = - m^2 \phi_{tt} + 2im \phi_{a}\phi_{at} + \phi_{a}\phi_{b}
\phi_{ab}, \quad M_2^* ,  \nonumber \\
S_{jk} (\phi_{ab},\phi_{ab}^*), \quad R^1_j = R_j (\theta_a,\phi_{ab}),
\quad R^2_j = R_j (\theta_a^*,\phi_{ab}),\quad R^3_j = R_j
(\phi_a+\phi_a^*,\phi_{ab}).  \label{yehorchenko:Galilei1 inv}
\end{gather}
Here $\theta_a = im \phi_{at} + \phi_{a} \phi_{ab}, \phi_{ab}$ are covariant
tensors for the Galilei algebra.

A functional basis of differential invariants for the Galilei algebra~(\ref
{yehorchenko:Galilei1}) extended by the dilation operator~(\ref
{yehorchenko:D}) and the projective operator~(\ref{yehorchenko:A}) may be
chosen as follows:
\begin{gather}
N_{1}e^{-2(\phi +\phi ^{\ast })},\quad \frac{N_{1}}{N_{1}^{\ast }},\quad
\frac{N_{2}}{N_{1}^{2}},\quad \frac{N_{2}^{\ast }}{(N_{1}^{\ast })^{2}}%
,\quad S_{jk}(\rho _{ab},\rho _{ab}^{\ast }),\quad R_{j}(\rho _{a},\rho
_{ab}),  \nonumber \\
R_{j}(\rho _{a}^{\ast },\rho _{ab}),\quad R_{j}(\phi _{a}+\phi _{a}^{\ast
},\rho _{ab})N_{1}^{-1},\quad (\phi _{aa}+\phi _{aa}^{\ast })N_{1}^{-1},
\label{yehorchenko:Galilei2 inv}
\end{gather}
where
\begin{equation}
N_{1}=M_{1}+\phi _{aa}=2im\phi _{t}+\phi _{aa}+\phi _{a}\phi _{a},\qquad
N_{2}=\frac{1}{n}\phi _{aa}N_{1}+\frac{\phi _{aa}^{2}}{2n}+M_{2}
\end{equation}
and the covariant tensors have the form
\[
\rho _{a}=\theta _{a}N_{1}^{-3/2},\qquad \rho _{ab}=\left( \phi _{ab}-\frac{%
\delta _{ab}}{n}\phi _{cc}\right) N_{1}^{-1}.
\]

An algorithm for construction of conditional differential invariants may be
derived directly from the Definition 4. Such invariants may be found by
means of the solution of the system~(\ref{yehorchenko:Gcond}),~(\ref
{yehorchenko:Xcond}).

We can construct conditional differential invariants of the Poincar\'{e}
algebra~(\ref{yehorchenko:poin}) and the projective operator~(\ref
{yehorchenko:A1}) solving the system
\[
{\mathop {A_1}\limits_{2}}F(\mathrm{Inv}_{P})=0,\qquad u_{0}+u_{3}+imu=0,
\]
where $\mathrm{Inv}_{P}$ are all differential invariants~(\ref
{yehorchenko:poin inv}) of the Poincar\'{e} algebra~(\ref{yehorchenko:poin}%
). Using the fact that the ansatz~(\ref{yehorchenko:ansatz}) is the general
solution of the additional condition~(\ref{yehorchenko:condition}), we can
directly substitute this ansatz into differential invariants~(\ref
{yehorchenko:poin inv}). To avoid cumbersome formulae here we did not list
expressions for all differential invariants from~(\ref{yehorchenko:poin inv}%
).

The expression $\Box u $ transforms into the following:
\[
\Box u = u (2im \phi_{\tau}+ \phi_{aa} + \phi_{a} \phi_{a}) ,
\]
where $N_1$ is an expression entering into expression for differential
invariants~(\ref{yehorchenko:Galilei1 inv}). Further we get
\begin{gather}
u_{\mu} u_{\mu}= u^2 ( 2im \phi_t + \phi_{a} \phi_{a}),  \nonumber \\
u_{\mu} u_{\nu} u_{\mu \nu} = u^3 (\phi_{a}\phi_{b} \phi_{ab}+ (\phi_{a}
\phi_{a})^2 - m^2 (\phi_{tt} +4 \phi_t^2) + \phi_{a}\phi_{b} \phi_{ab}+
(\phi_{a} \phi_{a})^2  \nonumber \\
\phantom{u_{\mu} u_{\nu} u_{\mu \nu} =}{} - m^2 (\phi_{tt} +4 \phi_t^2) +
2im \phi_{a}\phi_{at} + 4im \phi_t \phi_{a} \phi_{a}),
\label{yehorchenko:reduced basis}
\end{gather}
Substituting the ansatz~(\ref{yehorchenko:ansatz}) to all elements of the
fundamental basis~(\ref{yehorchenko:poin inv}) of second-order differential
invariants of the Poincar\'e algebra similarly to~(\ref{yehorchenko:reduced
basis}), we can obtain reduced basis of differential invariants, that may be
used for construction of all equations reducible by means of this ansatz. We
can give the following representation of the Poincar\'e invariants using
expressions $M_k$~(\ref{yehorchenko:Galilei1 inv}) and $N_k$~(\ref
{yehorchenko:Galilei2 inv}), where in the expressions for $M_k$, $N_k$ $%
(k=1,2)$ time variable is $\tau=x_0-x_3$:
\begin{gather}
\Box u = u N_1,\quad u_{\mu} u_{\mu} = u^2 M_1, \quad u_{\mu} u_{\nu} u_{\mu
\nu} = u^3 \left(M_2 + M_1^2\right),  \nonumber \\
u_{\mu \nu} u_{\mu \nu} = u^2 \left( 2M_2 + M_1^2 +
\phi_{ab}\phi_{ab}\right),  \nonumber \\
u_{\mu} u^*_{\mu}=\frac{uu^*}{2}\left(M_1+M^*_1-\left(\phi_{a}+\phi^*_{a}%
\right) \left(\phi_{a}+\phi^*_{a}\right)\right).
\end{gather}
Here $a$, $b$ take values from 1 to 2.

Whence
\begin{gather}
M_1 = u_{\mu} u_{\mu} u^{-2}, \quad \phi_{aa} = N_1 - M_1 = \frac{u \Box u -
u_{\mu} u_{\mu}}{u^2},  \nonumber \\
M_2 = u_{\mu} u_{\nu} u_{\mu \nu} u^{-3} - (u_{\mu} u_{\mu})^2 u^{-4}, \quad
N_1 = \frac{\Box u }{u},  \nonumber \\
N_2 = \frac{1}{n}\phi_{aa} N_1 + \frac{\phi_{aa}^2}{2n} + M_2 = u_{\mu}
u_{\nu} u_{\mu \nu} u^{-3} - (u_{\mu} u_{\mu})^2 u^{-4}  \nonumber \\
\qquad {}+\frac{1}{n} \frac{\Box u}{u} \frac{u \Box u - u_{\mu} u_{\mu}}{u^2}
+ \frac{1}{2 n} \frac{(u \Box u - u_{\mu} u_{\mu})^2}{u^4},  \nonumber \\
R_1 (\phi_{a} + \phi_{a}^*, \rho_{ab}) N_1^{-1}=(\phi_{a} +
\phi_{a}^*)(\phi_{a} + \phi_{a}^*) N_1^{-1}  \nonumber \\
\qquad =\left(N_1+N_1^*-{\frac{2 }{uu^*}} u_{\mu} u^*_{\mu}\right)N_1^{-1}=%
\frac{u^*\Box u +u\Box u^*-2u_{\mu} u^*_{\mu}}{u^*\Box u }.
\end{gather}

We construct Poincar\'e-invariant conditional differential invariants of the
projective operator~(\ref{yehorchenko:A1}) under the condition~(\ref
{yehorchenko:condition}) using differential invariants~(\ref
{yehorchenko:Galilei1 inv})
\begin{gather}
I_1 = N_1 e^{- 2(\phi + \phi^*)} = \frac{ \Box u}{u(uu^*)^2}, \quad I_2 =
\frac{N_1}{N_1^* } = \frac{ u^* \Box u}{u \Box u^*},  \nonumber \\
I_3 = \frac{N_2}{N_1^2 } = \left(u u_{\mu} u_{\nu} u_{\mu \nu} + \frac{3}{2 n%
} u^2 (\Box u)^2 + \left(\frac{1}{2n}-1\right) (u_{\mu} u_{\mu})^2 -\frac{2}{%
n} u \Box u(u_{\mu} u_{\mu})\right)\! {\left(u^2 (\Box u)^2\right)}^{-1},
\nonumber \\
I_4 =R_1 (\phi_{a} + \phi_{a}^*, \rho_{ab}) N_1^{-1}=\frac{u^*\Box u +u\Box
u^*-2u_{\mu} u^*_{\mu}}{u^*\Box u }.
\end{gather}

Whence, we may state that all equations of the form $F(I_1, I_2, I_3, I_4) =
0 $ are conditionally invariant with respect to the operator $A_1$~(\ref
{yehorchenko:A1}) with the additional condition~(\ref{yehorchenko:condition}%
).

Finding similar representations for all elements of the functional basis~(%
\ref{yehorchenko:Galilei1 inv}) of the second-order differential invariants
of the Galilei algebra~(\ref{yehorchenko:Galilei1}) extended by the dilation
operator~(\ref{yehorchenko:D}) and the projective operator~(\ref
{yehorchenko:A}), we can construct functional basis of conditional
differential operators. Such basis would allow to describe all
Poincar\'e-invariant equations for the scalar complex-valued functions that
are conditionally invariant under the operator $A_1$~(\ref{yehorchenko:A1}).

\section{Conclusion}

The procedure for finding conditional differential invariants outlined above
may be used for other cases when the additional condition~(\ref
{yehorchenko:Gcond}) has the general solution that may be used as ansatz,
and when a functional basis of the operator~(\ref{yehorchenko:Xcond}) in the
variables involved in such reduction is already known.

Besides finding new conditionally invariant equations, further developments
of the ideas presented in this paper may be description of all equations
reducible by means of a certain ansatz, and search of methods for
restoration of original equations from the reduced equations.

The symmetry of the nonlinear wave equation discussed in the paper may also
be interpreted as a hidden symmetry arising as symmetry of the reduced
equation. Thus the method described (construction of conditional
differential invariants) may also be used for description of equations
possessing hidden symmetry (see e.g. \cite{yehorchenko:Abraham}).

\end{document}